\documentclass[aip,prl,preprint,superscriptaddress]{revtex4-1}
\usepackage{graphicx}
\usepackage{units}
\usepackage{SIunits}
\usepackage{amsmath}
\usepackage{xcolor}
\usepackage{dcolumn}

\begin{document}

\title{On the impact of strain on the electronic properties of InAs/GaSb quantum well systems}

\author{L. Tiemann}\email{lars.tiemann@physik.uni-hamburg.de}
\affiliation{Solid State Physics Laboratory, ETH Zurich, 8093 Zurich, Switzerland}

\author{S. Mueller}\thanks{Contributed equally to this work}
\affiliation{Solid State Physics Laboratory, ETH Zurich, 8093 Zurich, Switzerland}
\author{Q.-S. Wu}
\affiliation{Theoretical Physics and Station Q Zurich, ETH Zurich, 8093 Zurich, Switzerland}
\author{T. Tschirky}
\affiliation{Solid State Physics Laboratory, ETH Zurich, 8093 Zurich, Switzerland}
\author{K. Ensslin}
\affiliation{Solid State Physics Laboratory, ETH Zurich, 8093 Zurich, Switzerland}
\author{W. Wegscheider}
\affiliation{Solid State Physics Laboratory, ETH Zurich, 8093 Zurich, Switzerland}
\author{M. Troyer}
\affiliation{Theoretical Physics and Station Q Zurich, ETH Zurich, 8093 Zurich, Switzerland}
\author{A. A. Soluyanov}
\affiliation{Theoretical Physics and Station Q Zurich, ETH Zurich, 8093 Zurich, Switzerland}
\affiliation{Department of Physics, St. Petersburg University, St. Petersburg, 199034, Russia}
\author{T. Ihn}
\affiliation{Solid State Physics Laboratory, ETH Zurich, 8093 Zurich, Switzerland}

\date{\today}

\begin{abstract}
Electron-hole hybridization in InAs/GaSb double quantum well structures leads to the formation of a mini band gap. We experimentally and theoretically studied the impact of strain on the transport properties of this material system. Thinned samples were mounted to piezo electric elements to exert strain along the [011] and [001] crystal directions. When the Fermi energy is tuned through the mini gap, a dramatic impact on the resistivity at the charge neutrality point is found which depends on the amount of applied external strain. In the electron and hole regimes, strain influences the Landau level structure. By analyzing the intrinsic strain from the epitaxial growth, the external strain from the piezo elements and combining our experimental results with numerical simulations of strained and unstrained quantum wells, we compellingly illustrate why the InAs/GaSb material system is regularly found to be semimetallic.
\end{abstract}


\maketitle

\section{Introduction}
Strain influences the band structure and the electronic properties of a semiconductor. The application of external strain can lift the valley degeneracy as observed in AlAs quantum well systems~\cite{Gunawan2006}, or change the spin decoherence times through the manipulation of the quadrupolar splitting in GaAs~\cite{Knotz2006, Chekhovich2015}. Epitaxial growth of alternating materials with different lattice constants results in intrinsically strained systems with modified electronic properties, which is, for example, exploited in modern high electron mobility field effect transistors. \newline
Here we study the influence of strain on the electronic properties of the InAs/GaSb double quantum well system. In a heterostructure consisting of an InAs and a GaSb quantum well (QW), a hybridization gap opens up when the respective QW widths are fine-tuned to bring the lowest conduction band state in the InAs QW energetically below the highest valence band state in the GaSb QW, enabeling hybridization of the electron and hole wave functions~\cite{Naveh1995, Quinn1996, Kroemer2004}. Experimentally, however, InAs/GaSb samples generally do not become insulating when the carrier density is tuned across the hybridization gap~\cite{Cooper1998, Knetz2010, Suzuki2013, Charpentier2013, Pal2015, Suzuki2015, Du2015} but exhibit a finite resistance instead. This semimetallic behavior has been attributed to a variety of mechanisms, such as to the anisotropy of the heavy hole band~\cite{Knetz2010}, or a perturbation of hybridization by scattering~\cite{Naveh1995, Naveh2001}.\newline
By using a combination of band calculations that incorporate strain and experiments with artificially strained InAs/GaSb systems mounted to piezo elements, we demonstrate that strain can not only be responsible for the semimetallic behavior but also strongly influences other transport properties of this material system. More specifically, our data show that InAs/GaSb can generally be a semimetal with a finite conductance at the charge neutrality point due to intrinsic strain originating from the lattice mismatch between InAs and GaSb and the buffer material underneath. We use external strain applied via a piezo ceramic to further modify the electronic properties, including the charge carrier density and observe changes in magneto-transport. 

\newpage

\section{SAMPLE DETAILS}
\begin{figure}[!h]
\includegraphics[width=16cm]{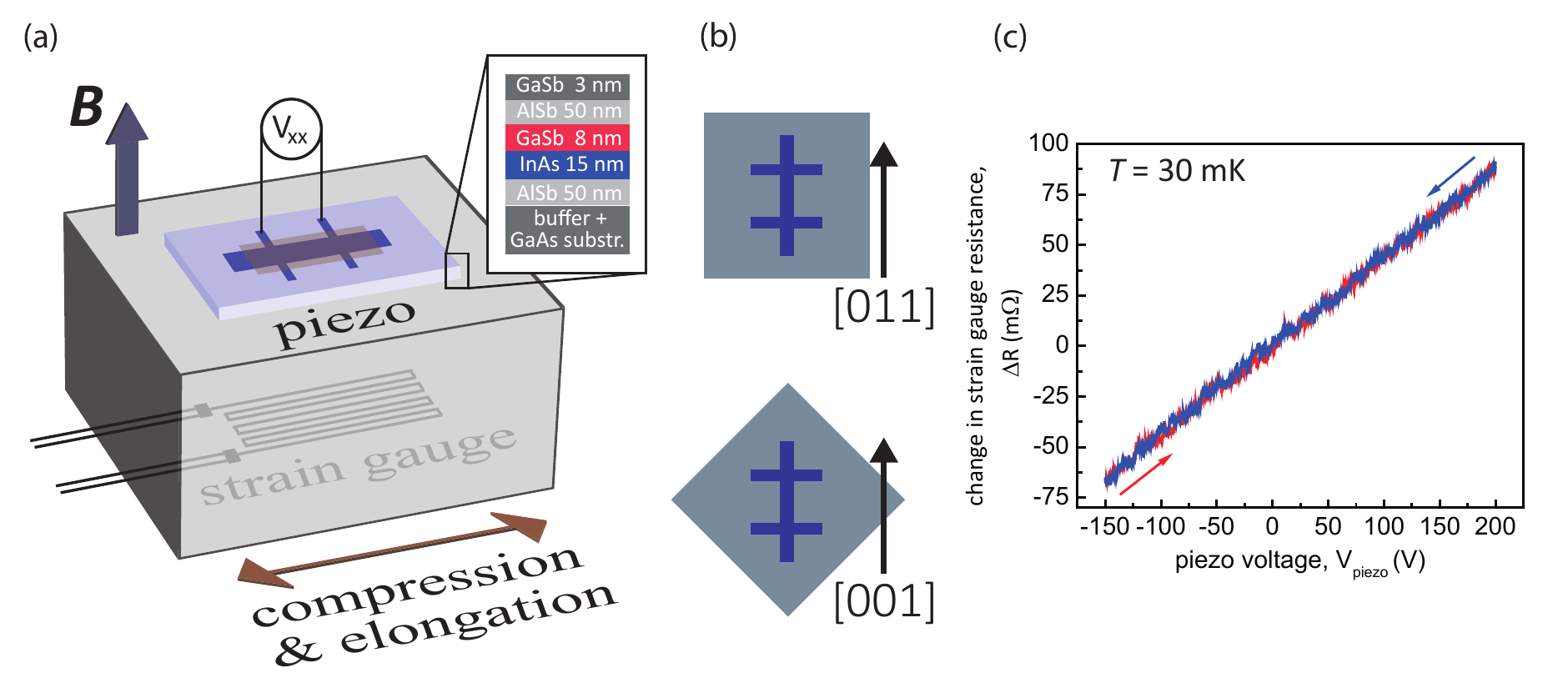}
\caption{\label{fig1}{{\bf (a)} Schematics of the piezo stack with the strain gauge underneath and the sample mounted on top. The Hall bars are always aligned along the piezo’s direction of compression (for negative voltages) and elongation (for positive voltages). The inset shows the heterostructure. {\bf (b)} The Hall bars were either patterned to be orientated along the [011] or the [001] direction. Shown is a top view of the sample’s (100) surface. {\bf (c)} Demonstration of the perfect linearity of our piezo ceramics, exemplarily showing the strain gauge resistance change at a temperature of 30 mK for an upsweep (red curve) and downsweep (blue curve) of the piezo voltage. }}
\end{figure}

We have investigated four samples from two different wafers to guarantee the reliability of our data and the validity of our conclusions. Samples 1,2 and 4 were fabricated from wafer \#1, whereas sample 3 was fabricated from wafer \#2. The heterostructures [see inset of Fig.\;\ref{fig1}(a)] are nominally the same for both wafers and were grown on GaAs substrates in direct succession. The heterostructures consist of an 8 nm GaSb layer on top of 15 nm InAs~\cite{Charpentier2013}. Beneath, 50 nm AlSb, a 50 nm superlattice of AlSb/GaSb, 500 nm GaSb, a layer of Al$_{x}$Ga$_{1-x}$Sb ($x$ = 0.65) and additional buffer materials are used to ease the lattice mismatch with the GaAs substrate. For the sample preparation, we followed the process detailed in Ref.~\onlinecite{Pal2015} and in Appendix \ref{app_0}.

In our experiments we use piezo-ceramic stacks to apply external strain to the samples. To maximize the amount of external strain that is exerted, the samples are thinned from the back in a chemical-mechanical bromine methanol etching process to 50-70 $\mathrm{\mu}$m. As a precaution, to screen electric stray fields from these piezos, a 100 nm thick layer of Cr/Au was evaporated on the back of each sample. This screening layer is grounded during the experiments. Each piezo stack has a nickel-chromium strain gauge glued to one of its insulated sides. The resistance of the gauge changes in response to length variations of the piezo. Figure\;\ref{fig1}(a) shows a simplified schematic.

\section{STRAIN MAGNITUDES AND DIRECTIONS}

The physics of the InAs/GaSb double QW system is governed by the hybridization of the lowest subband in the InAs QW and the highest subband in the GaSb QW, whose energies are determined by the respective well widths. When the piezo elongates (contracts) the sample along [001] or [011] [Figs.\;\ref{fig1}(a) \& (b)], it will shrink (expand) in the perpendicular growth direction, i.e., along [100]. This will slightly modulate the quantum well widths, their energy levels and the resulting hybridization.\newline

In addition to the external strain that we apply with our piezos, the double quantum well system is subject to intrinsic biaxial strain due to the epitaxial pseudomorphic growth of different semiconductor materials with different bulk lattice constants (AlSb 6.1355 \AA, GaSb 6.0959 \AA, InAs 6.0583 \AA\;at 300 K). For simplicity we denote both intrinsic and external strain as $\mathrm{\varepsilon}$. In Appendix \ref{app_A}, we have estimated the intrinsic strain for the ideal situation of a complete relaxation of all materials to the GaSb lattice constant. In this case, only the InAs quantum well is strained ($\mathrm{\varepsilon_{InAs}}\sim$ 54 $\times$ 10$^{-4}$), whereas the GaSb quantum well is unstrained $(\mathrm{\varepsilon_{GaSb}}=0)$. The tunable external strain, which we estimate to be in the range $-1.3 \times 10^{-4} <\mathrm{\varepsilon} < +1.7 \times  10^{-4}$ ($\Delta\mathrm{\varepsilon}\sim 3 \times 10^{-4}$) at low temperatures, is added to the intrinsic strain and will modify or even enhance the lattice distortion due to the lattice mismatch at the InAs/GaSb interface~\cite{Hu2016}. Please note that the piezo adds strain in a non-trivial way, e.g., when the piezo elongates in one direction, it will shrink in its perpendicular direction just like the sample which is glued to its surface. That means that the piezo strain is not uniaxial. These effects have been carefully included into our band structure calculations. \newline

We will demonstrate below that despite the smallness of the external strain, the piezo elements can dramatically alter the resistivity at the charge neutrality point and other transport properties.

\section{MEASUREMENTS}

The samples were cooled down with a 10 M$\mathrm{\Omega}$ resistor between the piezo connectors to avoid thermal voltage build-up and a possible damage of the piezo stacks. At room temperature and prior to the measurements, the piezo voltages were swept several times to test their performance by monitoring the strain gauge. While piezo mechanical stacks show hysteresis and non-linear behavior at room temperature, at low temperatures these effects vanish and linear bipolar operation over large voltage ranges is possible \cite{Shayegan2003}.

Figure\;\ref{fig1} (c) illustrates the 4-terminal resistances of a strain gauge versus piezo voltage at 30 mK. We would like to stress that the strain gauge can obviously only detect relative changes due to the elongation or contraction of the piezo stack; it is insensitive to the intrinsic strain resulting from the lattice mismatch in the sample.

\begin{figure}[!h]
\includegraphics[width=16cm]{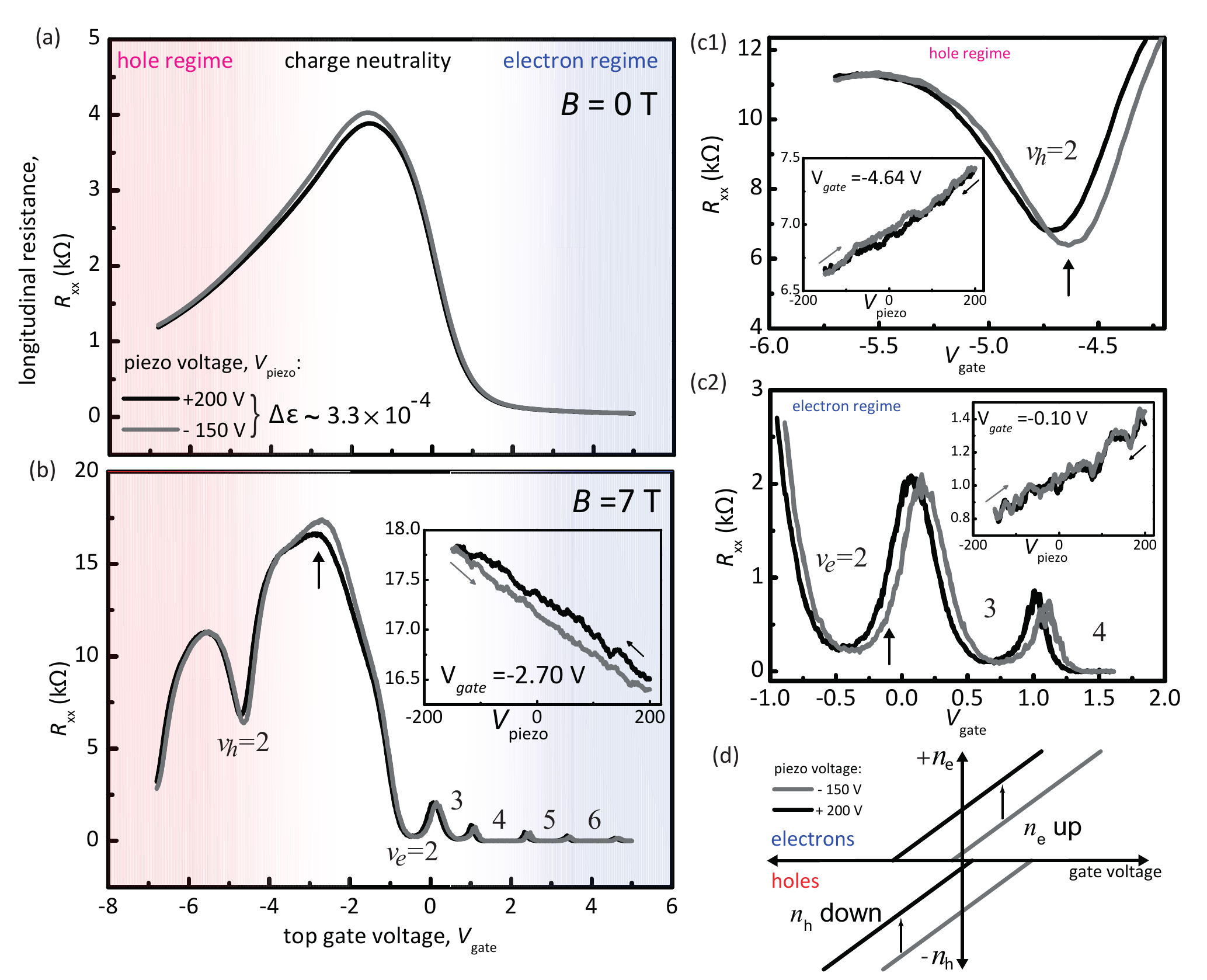}
\caption{\label{fig2}{The longitudinal resistance, $R_{xx}=V_{xx}/I$, versus gate voltage at two exemplary magnetic fields at $B$ = 0 T {\bf (a)} and $B$ = 7 T {\bf (b)}  at 20 mK for the maximal piezo voltages, i.e., for maximal compressive strain and maximal tensile strain (sample 4, Hall bar aligned in the [011] direction). For $\Delta V_{\mathrm{piezo}}$ = 350 V, we estimated the strain range to be of the order of $3.3 \times 10^{-4}$ (see Appendix\;\ref{app_A}). Each sweep was repeated twice to demonstrate the stability (Appendix\;\ref{app_B}). {\bf (c1)} \& {\bf (c2)} Blow-ups of the shifts in the Landau level positions. The insets show $R_{xx}$ vs. piezo voltage at a fixed gate voltage (indicated by the nearby arrow).{\bf (d)} Illustration on how the carrier densities (filling factors) respond to the strain. Changing the piezo voltage from -150 V to +200 V at a fixed gate voltage results in an increase in the electron density but a decrease in the hole density.}}
\end{figure}

Figure\;\ref{fig2} (a) shows the longitudinal resistances, $R_{xx}$, versus the top gate voltage, $V_{\mathrm{gate}}$, at zero magnetic field, $B$, and Fig.\;\ref{fig2} (b) at 7 tesla (T) for the two maximal piezo voltages that were applied, i.e., for the maximal elongation and compression of the piezo. In this sample (sample 4) the Hall bar is oriented along the [011] direction. For each piezo voltage, we have performed two consecutive gate voltage sweeps in the same direction and observe perfect reproducibility (see Appendix\;\ref{app_B}).

The variation of the top gate voltage changes the carrier concentration and carrier type by shifting the Fermi energy from the conduction band, through the hybridization gap into the valence band. In Figure\;\ref{fig2} (b) at finite magnetic fields, the density of states splits into dispersionless Landau levels and integer quantum Hall effects emerge~\cite{Klitzing1980}. Varying the top gate voltage at constant magnetic field of 7 T, empties or fills these Landau levels while their degeneracy remains constant. For positive voltages, $V_{\mathrm{gate}} > 0$ V, at $B$ = 7 T the well-known Shubnikov-de-Haas oscillations indicate the integer quantum Hall effect, which arises from the aforementioned Landau level quantization for the electrons in the InAs QW. The resistance maximum around $V_{\mathrm{gate}}\sim$ -2.5 V is interpreted as the charge neutrality point, where the Fermi energy is expected to reside in the hybridization gap. However, we do not find the sample to become insulating. For $V_{\mathrm{gate}} < -4$ V, the onset of the integer quantum Hall effect of holes is seen, which is less pronounced than for electrons due to their larger effective mass. 

\newpage

We find two distinct effects that result from changes in the in-plane strain: (1) a dependence of the peak height at the charge neutrality point [Figs.\;\ref{fig2} (a) \& (b)], and (2) shifts in the position of the resistivity minima [blow-ups for $B$ = 7 T in Figs.\;\ref{fig2} (c) indicate these shifts]. The impact on the resistance at the charge neutrality point is directly seen in Figs.\;\ref{fig2} (a) \& (b), while the shifts in the minima is only seen upon closer inspection in Figs.\;\ref{fig2} (c1) \& (c2). The insets in Figs.\;\ref{fig2} (b) and (c) demonstrate that these effects scale linearly with the piezo voltage when the gate voltage is fixed to values marked by the arrows in Fig.\;\ref{fig2} (b) and Figs.\;\ref{fig2} (c). At the charge neutrality point, we observe some hysteretic behavior between up- and downsweeps of the piezo voltage. These two effects, i.e., the variations of the resistance near the charge neutrality point and the aforementioned shift, appear consistently within the range of magnetic fields between 0 T and 12 T and in all samples, however, at magnet fields exceeding 9 T, strong "noise" is overlaid to resistance at the charge neutrality point. Further examples are shown in Appendix\;\ref{app_B}. Both effects results from the band distortion upon strain which we will address in the next section.  \newline

To ensure that these effects are really the result of strain and not an artefact arising from stray electric fields from the piezo for example, a fifth sample was prepared, which rested on the piezo stack but was only glued to it at one of its edges. In this test sample, which was not influenced by strain, we did not observe any effect on the transport properties of the sample when the piezo was driven with large positive or negative voltages. This demonstrates that the observations described above are purely strain-related.\newline

The significance of the resistivity minima shift in gate voltage is illustrated in Fig.\;\ref{fig2} (d). Generally, sweeping the gate voltage results in a linear response of the carrier density. The piezo-strain induces variations of the (intrinsic) carrier density in the sample, which in turn results in a shift of the density-voltage curve, i.e., under strain different gate voltages are needed to access the same filling factors. For electrons, the density increases upon sample elongation and decreases upon its compression, while for the holes, we observe the opposite behavior. From these shifts and from additional measurements of the Hall resistance we have extracted the change in carrier density arising from strain. 

\newpage

\begin{figure}[!h]
\includegraphics[width=12cm]{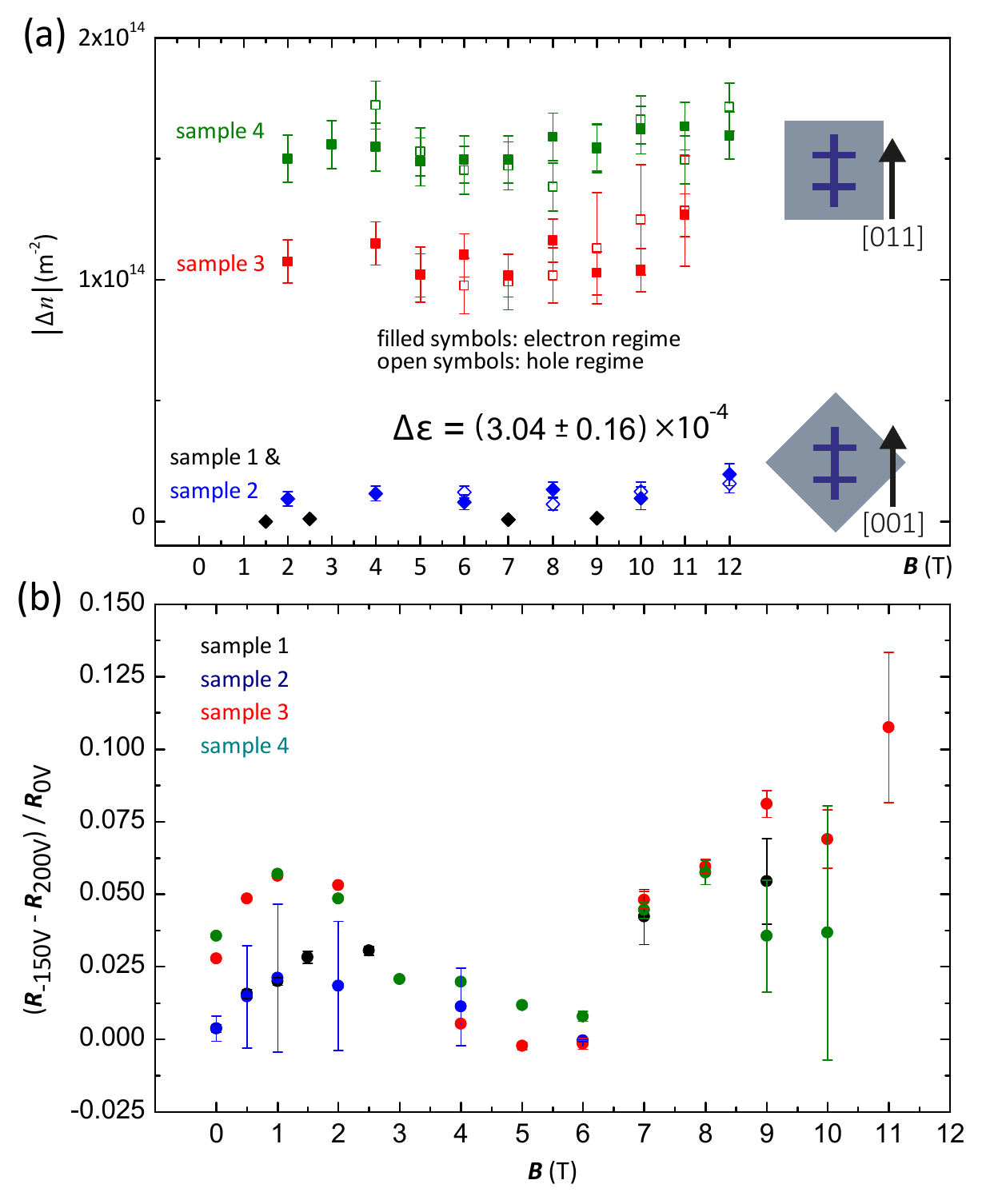}
\caption{\label{fig3}{{\bf (a)} Absolute density change for electrons and holes between maximal tensile and compressive strain versus magnetic field. $\Delta n$ is positive (negative) for electrons (holes). The maximal external strain for each sample, $\Delta\varepsilon = (3.04 \pm 0.16)\times 10^{-4}$ (a dimensionless number), varies only slightly between samples. Each color corresponds to one sample. Samples 1 \& 2 (3 \& 4) with the Hall bar aligned in the [001] ([011]) direction. {\bf (b)} $R_{\mathrm{-150V}} - R_{\mathrm{+200V}} / R_{\mathrm{0V}}$, i.e, the normalized change in the peak resistance close to the charge neutrality point between maximal tensile strain and maximal compressive strain. The peak resistance slightly shifts with the magnetic field, $B$. At large $B$, strong fluctuations in the resistance appear (see Appendix\;\ref{app_B}).}}
\end{figure}

Figure\;\ref{fig3} summarizes the measurements performed on all four samples, including measurements on two Hall bars that are aligned along the [001] direction. The density change, $\Delta n$, only depends on the piezo strain and the sample orientation, with samples aligned along [011] displaying the strongest response to strain. As expected, $\Delta n$ is independent of the magnetic field [Figure\;\ref{fig3} (a)]. Within our measurement uncertainty, we find that electron and hole densities change by approximately the same absolute value. The [011] direction of the zincblende unit cell is not a high symmetry axis, which makes it more susceptible to symmetry breaking strain; a behavior we observe in the experiments. This susceptibility to strain is reflected in the larger number of terms in the tight-binding model that we use to numerically calculate the band structures below.\newline

Figure\;\ref{fig3} (b) shows the change in ''peak'' resistance around the charge neutrality point between maximal tensile and maximal compressive strain, normalized to its resistance when no external strain was applied. Although the density does not depend on the magnetic field, we found that the top gate voltage corresponding to the peak resistance slightly shifts with $B$, which results in a very different magnetic field dependence with respect to Fig.\;\ref{fig3} (a). We have no consistent picture to explain this behavior. However, since Coulomb energies scale with magnetic field, it is feasible that charges in this low carrier density regime will be very susceptible to changes in $B$.\newline

To understand and interpret these results, we have performed numerical band structure simulations for a strained InAs/GaSb system. 

\newpage

\section{BAND STRUCTURE SIMULATIONS AND ANALYSIS\label{sec_band_calc}}
Numerical simulations of strained and unstrained quantum wells were performed using symmetrized Wannier-based tight-binding (TB) models. First, symmetrized TB models with long-range hoppings are constructed for bulk InAs and GaSb\cite{Soluyanov-PRB15}. To approximate the exchange correlations, a first-principles simulation of a bulk material is carried out using HSE hybrid functional without spin-orbit coupling (SOC). The resultant spectrum within a specially chosen energy window is then projected onto the In (Ga) $s$- and $p$-states and As (Sb) $p$-states to construct a 7$\times$7 Wannier-based TB model~\cite{Souza-PRB06, Wannier90}. This model is symmetrized to satisfy all the spatial symmetries of the zincblende $T_d$ crystal structure. Local SOC is added separately to this model in such a way that the spectrum of the resultant TB model coincides with a HSE simulation with SOC~\cite{Soluyanov-PRB15}. We carried out the HSE hybrid functional~\cite{HSE06} simulations using the VASP software package~\cite{VASP}. The screening parameters $\mu$ used in the HSE method were adjusted to 0.2 and 0.15 for InAs and GaSb, respectively, such that the experimental band gap is fitted~\cite{Kim10}. The PBE functional for exchange-correlation~\cite{PBE} and PAW~\cite{PAW1, PAW2} pseudopotantials were employed, with SOC implemented in the pseudopotentials. The energy cut-off was $380$~eV for both InAs and GaSb. We used a $6\times6\times6$ $\Gamma$-centered $k$-mesh and Gaussian smearing of width $50\;\mathrm{meV}$ for Brillouin zone integrations. To simulate the interface between the two quantum well materials, we interpolated~\cite{Wu-prep16} the hopping parameters. This procedure provides reasonable agreement with experiment for InAs/GaSb quantum wells~\cite{Wu-prep16}.

The underlying strain model maps the crystal structure of an idealized InAs/GaSb system, in which the two quantum wells were grown pseudomorphically, with InAs having adopted the lattice constant of GaSb. Comparison of the experimental data with theory will later show that in our samples also the GaSb quantum well is subject to a small amount of growth-related strain. Additive external strain is incorporated into our model by calculating the response of the primitive lattice vectors to (uniaxial) strain in the [001] and [011] directions and by defining the strain tensor with the elastic stiffness constants for InAs and GaSb. Symmetrized TB models were then constructed for the strained bulk materials first, following the procedure described above. After interpolation to simulate the interface, the resultant TB models are used to obtain the band structure and DOS for various values of strain for the InAs/GaSb quantum wells. 

\begin{figure}[!h]
\includegraphics[width=16cm]{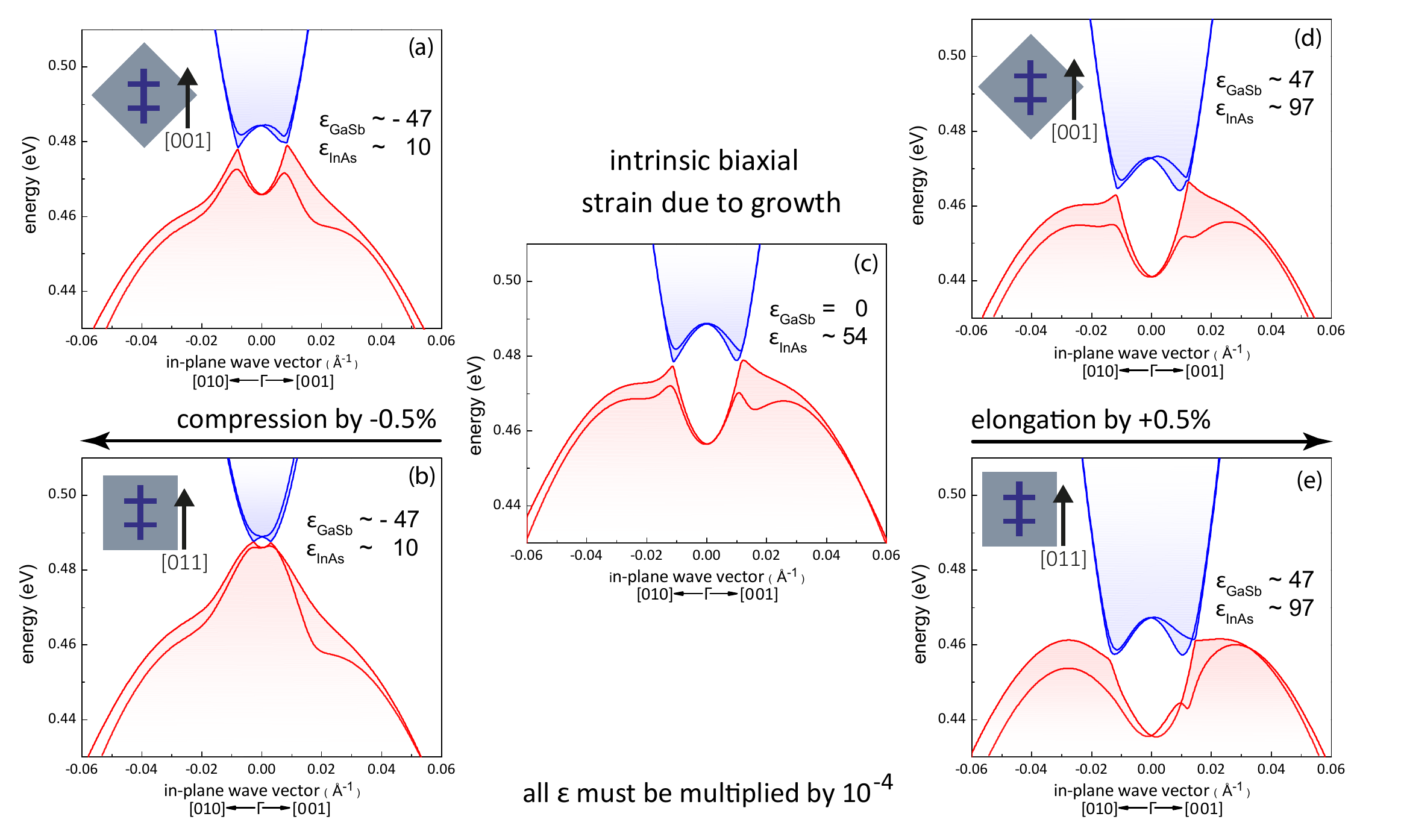}
\caption{\label{fig4}{{\bf (c)} Band structure without external strain. {\bf (a) \& (b)} Under compression by -0.5\% along [001] and [011]. {\bf (d) \& (e)} Under elongation by +0.5\% along [001] and [011]. Note that the in-plane wave vector is only displayed with its two in-plane directions.}}
\end{figure}

We have performed TB band structure calculations for several amounts of compression and elongation along [001] and [011] by the piezo and for the same heterostructure as in our experiments. Figure\;\ref{fig4} shows a selection of exemplary results. Our band structure calculations use the C$_{\mathrm{2v}}$-symmetry for the zincblende structure, which does not display radial symmetry. For that reason, the band structure in the two orthogonal directions in $k$-space will differ. The center figure (c), represent the growth condition without external strain. Figures\;\ref{fig4} (a) and (b) show the band structures when the unit cell is compressed by -0.5\% along [011] or [001], while Figs.\;\ref{fig4} (d) and (e) show the band structures for a corresponding elongation by +0.5\%. Because the intrinsic strain in the InAs layer surpasses the external strain, $\varepsilon_{\mathrm{InAs}}$ is always positive, whereas for GaSb, $\varepsilon_{\mathrm{GaSb}}$ can be both positive or negative. Even without any detailed analysis, the strong band distortion for only 0.5\% strain implies a dramatic impact on the transport properties of this material system.
 Figure.\;\ref{fig4} (c) implies that an 8 nm GaSb / 15 nm InAs double quantum well system~\cite{Knetz2010, Charpentier2013} is essentially semimetallic, even when no external strain is applied. For future experiments, we suggest an 8 nm GaSb / 12 nm InAs double quantum well system instead which by contrast displays a well-developed mini-gap (Appendix\;\ref{app_D}).\newline

 From the band structures, we calculated the respective density of states (DOS), from which, in turn, we extracted the electron and hole densities. We have outlined these calculations in Appendix\;\ref{app_C}. Figures\;\ref{fig5} (a) and (b) illustrate the theoretically expected densities as a function of energy for different values of strain along [001] and [011]. At energies exceeding $\sim$ 0.47 eV, the InAs QW is populated by electrons (while the GaSb QW is depleted) and at lower energies the GaSb QW is populated by holes (while the InAs is depleted). Figures\;\ref{fig5} (c) and (d) represent cross sections at two fixed energies, i.e., they show how electron and hole densities change when strain is applied (at a fixed energy). While we calculated the band structures for strain ranging from -1\% to +1\%, the strain we can exert with our piezo electric elements, $\Delta\varepsilon$, is comparatively small, as indicated by the magenta-colored vertical bars.
 
 Our band structure calculations and the subsequent analysis exhibit the same trends as the experimental data, which demonstrates the overall reliability of our idealized model. In particular, the stronger response shown for the [011] crystal direction is reproduced. For example, Fig.\;\ref{fig5} (c) shows the electron regime at 0.52 eV, where we find that strain along [011] yield a stronger increase in the electron density, or a much larger shift in the SdH minima, respectively. This is indeed what we observe experimentally. The picture is ambiguous for the holes since it is much more dependent on the energy at which we scrutinize the system. Experimentally, we are only able to observe the onset of hole transport near the CNP, where electrons may still co-exist. Figure\;\ref{fig5} (d) depicts the hole density as a function of strain at 0.46 eV. For strain $> 0 \%$, the model anticipates a decreasing hole density with a stronger response for strain along [011]. Also this is in agreement with experiment. Our piezo generates a $\sim$ 1.5 - 1.75 $\%$ electron density increase for [011], while theoretically at a Fermi energy of 0.52 eV we find 3 $\%$. For the [001]-direction, the experimental values are $\sim$ 0.15 $\%$, while theory predicts 1.4 $\%$ (at $\Delta E_F$ = 0.52 eV). For much larger values of strain, which we cannot generate with our piezos, the model anticipates that the hole density should change more rapidly with $\varepsilon$ than the electron density. The experiments, however, do not reproduce the non-monotonic behavior of Fig.\;\ref{fig5} (d) around 0\% strain. Experimentally, we had observed a linear dependence in the hole regime on the piezo voltage instead [Figure\;\ref{fig2} (c1)]. We believe that the reason for this discrepancy lies in the AlSb-rich buffer material, whose lattice constant exceeds that of GaSb so that it already adds additional (intrinsic) strain to the InAs and GaSb QWs. Thus, when we apply external strain to our samples, we operate within the range indicated by the error bar (based on an estimate using the AlSb lattice constant, similar to Appendix\;\ref{app_A}).\newline

\begin{figure}[!h]
\includegraphics[width=16cm]{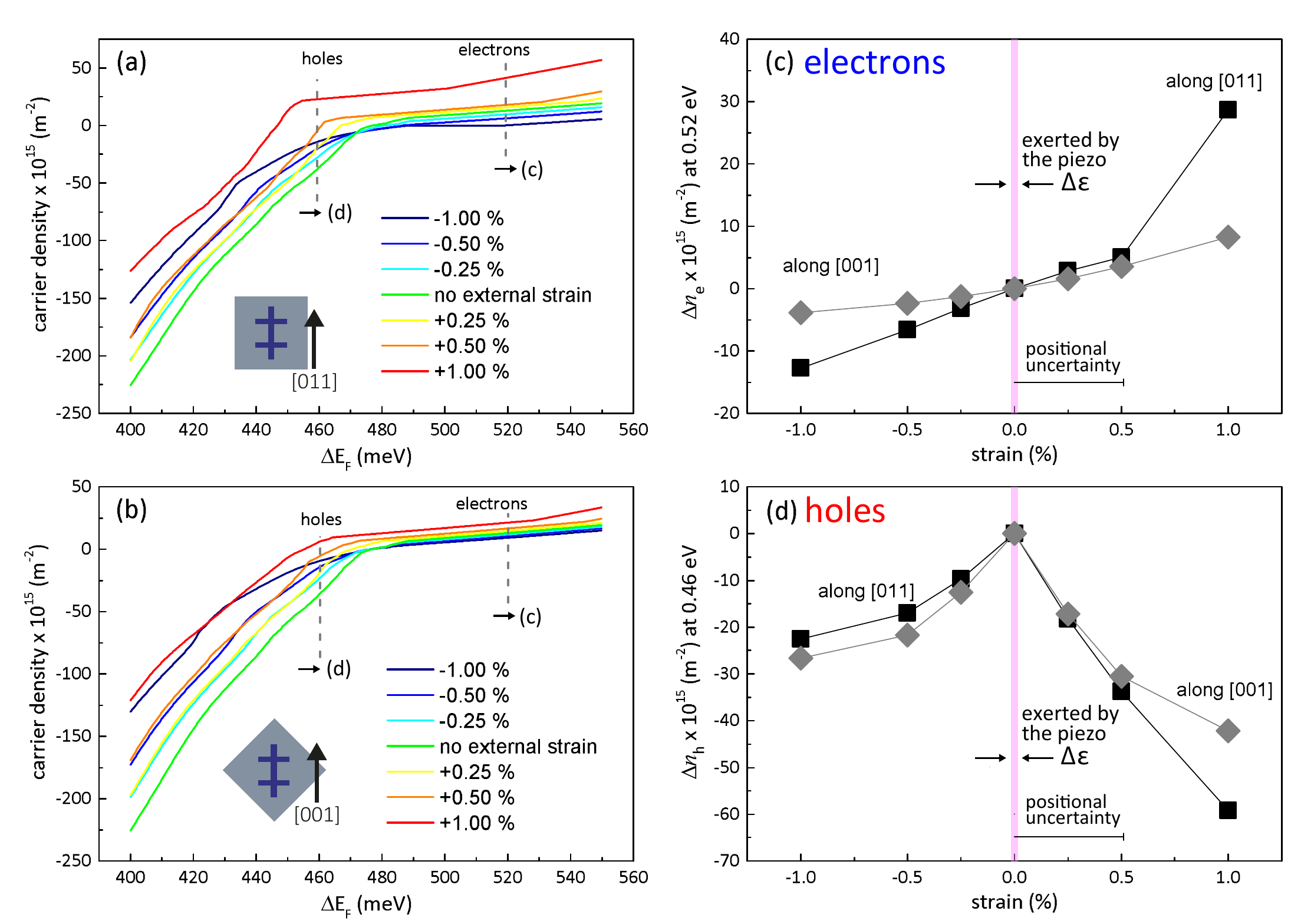}
\caption{\label{fig5}{{\bf (a) \& (b)} Electron and hole densities as a function of the Fermi energy for all calculated strain values and directions. The dashed lines at 0.46 eV and 0.52 eV are cross sections whose results are shown in subfigures (c) and (d). {\bf (c)}  Electron density change, $\Delta n_e$, as a function of strain along [011] and [001]. {\bf (d)}  Hole density change, $\Delta n_h$,  as a function of strain along [011] and [001].  The magenta-colored vertical bar indicates the amount of strain, $\Delta\varepsilon$, which is actually exerted by the piezo; its position is subject to an uncertainty (see main text for details).}}
\end{figure}

\section{INTERPRETATION AND CONCLUSION}
Our calculations have illustrated the strong impact of strain on the band edge positions. Although the strain, which is induced by the piezo elements, is small as compared to the intrinsic strain from the pseudomorphic growth, we have shown that it still has considerable effect on the resistance in the vicinity of the charge neutrality point. We therefore conclude that strain is a dominating parameter in this material system and plays a major role for the semimetallic behavior regularly observed in the InAs/GaSb quantum well system. In this context, Zakharova~\textit{et al.}~\cite{Zakharova2002} have discussed the possibility of a strain-induced semimetal-semiconductor phase transitions in InAs/GaSb quantum wells which would emerge on InAs substrates from the resulting lattice mismatch.

In addition to inducing the semimetallic behavior, lattice-mismatched strain has been identified theoretically to influence the Landau level structure~\cite{Zakharova2004}. We indeed found that we can study the changes in the positions of the SdH minima resulting from variations in the carrier density which qualitatively agree with those density changes extracted from our band simulations. The density change shows a clear direction-dependence in favor of the [011] direction. Hall bars oriented along [001] do not exhibit large density variations but are still semimetallic because the intrinsic lattice mismatch is strong and ubiquitous. 

We believe that these experiments have compellingly illustrated that strain is an important element in this material system. Strain engineering is an additional tool to improve the insulating bulk properties in InAs/GaSb. For fine-tuned heterostructures, i.e., when the epitaxial growth produces minimally strained layers, small additional external strain may allow an in-situ control of a semimetallic – insulating transition.\newline

After the completion of this manuscript, we became aware of two related works (Refs. \onlinecite{DuA2016, MurakiA2016}), one of them using a series of pseudomorphically-strained InAs/In$_x$Ga$_{1-x}$Sb samples with different values for the composition factor $x$.\newline

\newpage

\begin{acknowledgments}
The authors acknowledge financial support by the Swiss National Science Foundation (SNF) and the NCCR QSIT (National Competence Center in Research - Quantum Science and Technology). A.A.S., Q.S.W., and M.T. were supported by Microsoft Research. We would also like to acknowledge Roland Winkler for sharing his insight on strain in semiconductors.
\end{acknowledgments}

\newpage

\begin{appendix}

\section[1]{\label{app_0} SAMPLE PREPARATION}

The sample preparation begins with a standard optical lithography process and a chemical etch to define a Hall bar structure. Ohmic contacts were obtained by optical lithography and the evaporation of Au, Ge and Ni. The devices were then passivated with 200 nm of Silicon Nitride, Si$_{3}$N$_{4}$, deposited by Plasma Enhanced Chemical Vapor Deposition (PECVD) at a temperature that also acts as an annealing step for the Ohmic contacts. Si$_{3}$N$_{4}$ serves as the dielectric for a Ti/Au top gate, produced by evaporation through a shadow mask. The top gate is used to tune the system from the electron transport regime in the conduction band (InAs) through charge neutrality point into hole transport in the valence band (GaSb).

The sample surfaces were then protected with a thick layer of optical photo resist and thinned from the back side in a chemical-mechanical bromine methanol etching process to 50-70 $\mathrm{\mu}$m. To screen the large electric fields from these piezos which are driven with up to 200 Volts, a 100 nm thick layer of Cr/Au was evaporated on the back side of each sample. The samples were then transferred to a ceramic chip carrier with a removable base plate for wire bonding.

We use lead-zirconium-titanate piezo-ceramic stacks which are specifically designed for the use in low temperature and vacuum conditions (Piezomechanik GmbH Munich, Germany). Its piezoelectric coefficient d$_{33}$, is positive, i.e., positive piezo voltages will lead to an elongation while negative voltages lead to its contraction along the poling direction. Each piezo stack has a nickel-chromium strain gauge (Vishay Precision Group, linear pattern type) glued to one of its insulated sides. The resistance of the gauge changes in response to lengths variations of the piezo. The glue is a two-component epoxy with a thermal characteristic specified down to 4 Kelvin (M-BOND 610, Vishay Precision Group). The rear face of the piezo stack is mounted to a sample holder with the strain gauge facing down. After removing the base plate of the chip carrier, the sample is freely hanging on its bonding wires, and it can then be lowered and glued to the piezo stack. 

\newpage

\section[2]{\label{app_A} STRAIN ESTIMATES}
We estimate the intrinsic strain $\varepsilon$ due to the pseudomorphic growth of materials with different lattice constants following Ref.\;\onlinecite{Cebulla1988}, where:

\begin{equation*}
\varepsilon=\frac{a_\parallel-a_1}{a_\parallel}
\end{equation*}
with 
\begin{equation*}
a_\parallel=\frac{a_1d_1+a_2d_2}{d_1+d_2},
\end{equation*}

while ignoring the elastic compliance constants. Here $a_1$ is the lattice constant of the material 1 to be grown on to the material 2 with the lattice constant $a_2$. The constants $d_1$  and $d_2$  are the corresponding layer thicknesses. The lattice constants we use are AlSb 6.1355 \AA, GaSb 6.0959 \AA, and InAs 6.0583 \AA.  The double quantum well structure consists of an 8 nm GaSb layer on top of 15 nm InAs. Beneath, 50 nm AlSb, a 50 nm superlattice of AlSb/GaSb, 500 nm GaSb, a 1100 nm thick layer of Al$_{x}$Ga$_{1-x}$Sb (x = 0.65, lattice constant ca. 6.12 \AA) and additional buffer materials are used to ease the lattice mismatch with the GaAs substrate.

We assume that InAs grows pseudomorphically on the first two buffer layers (50 nm AlSb and 50 nm total thickness of a AlSb/GaSb superlattice), which are thin and have adapted to the lattice constant of the GaSb underneath, i.e., also InAs is intrinsically strained. For 15 nm InAs on a 100 nm thick layer with a lattice constant of 6.0959 \AA\;this results in a strain of approximately $\varepsilon\sim 54 \times 10^{-4}$ (lattice constant mismatch 0.62\%). If also the InAs layer has adopted the lattice constant of GaSb, then the 8 nm GaSb quantum well will grow unstrained (no strain). 

The estimate for the external strain exerted via the piezo elements is based on the manufacturer’s specifications of strain/volt efficiency, the maximal stroke (change in length) at room temperature. At room temperature, the maximal strain from the piezo is ca.  $-3.1 \times 10^{-4} <\varepsilon< +15.5 \times 10^{-4}$ (based on the specifications by the manufacturer). At low temperatures, the strain/volt efficiency drops to a value not specified by the manufacturer. At the same time, however, the piezo allows bipolar operations over much larger voltages to compensate for the reduction in strain/volt efficiency. Since we measure the strain gauge resistance at room temperature and low temperature, we can compare both measurements to find the absolute reduction in strain/volt efficiency, while taking into account the gauge's gauge factor and thermal output, i.e., the correction due to thermal changes~\cite{vishaypg2016}. We estimate the strain range exerted by the piezos at 20 mK to be of the order of \newline

\noindent$-1.07 \times 10^{-4}< \varepsilon < +1.83 \times 10^{-4}\;(\Delta\varepsilon = 2.90 \times 10^{-4})$ for sample 1\newline
$-1.32 \times 10^{-4}< \varepsilon < +1.75 \times 10^{-4}\; (\Delta\varepsilon = 3.07 \times 10^{-4})$ for sample 2\newline
$-1.41 \times 10^{-4}< \varepsilon < +1.89 \times 10^{-4}\; (\Delta\varepsilon = 3.30 \times 10^{-4})$ for sample 3\newline
$-1.25 \times 10^{-4} < \varepsilon < +1.65 \times 10^{-4}\; (\Delta\varepsilon = 2.90 \times 10^{-4})$ for sample 4\newline

Our values are in good agreement with Ref.~\onlinecite{Shayegan2003}, but we want to stress that strain measured by the gauges are influenced by the thermal properties of the glue, which differs from the one used in the cited reference. 

\newpage

\section[3]{\label{app_B} ADDITIONAL MAGNETO-TRANSPORT TRACES}

\begin{figure}[h]
\includegraphics[width=8cm]{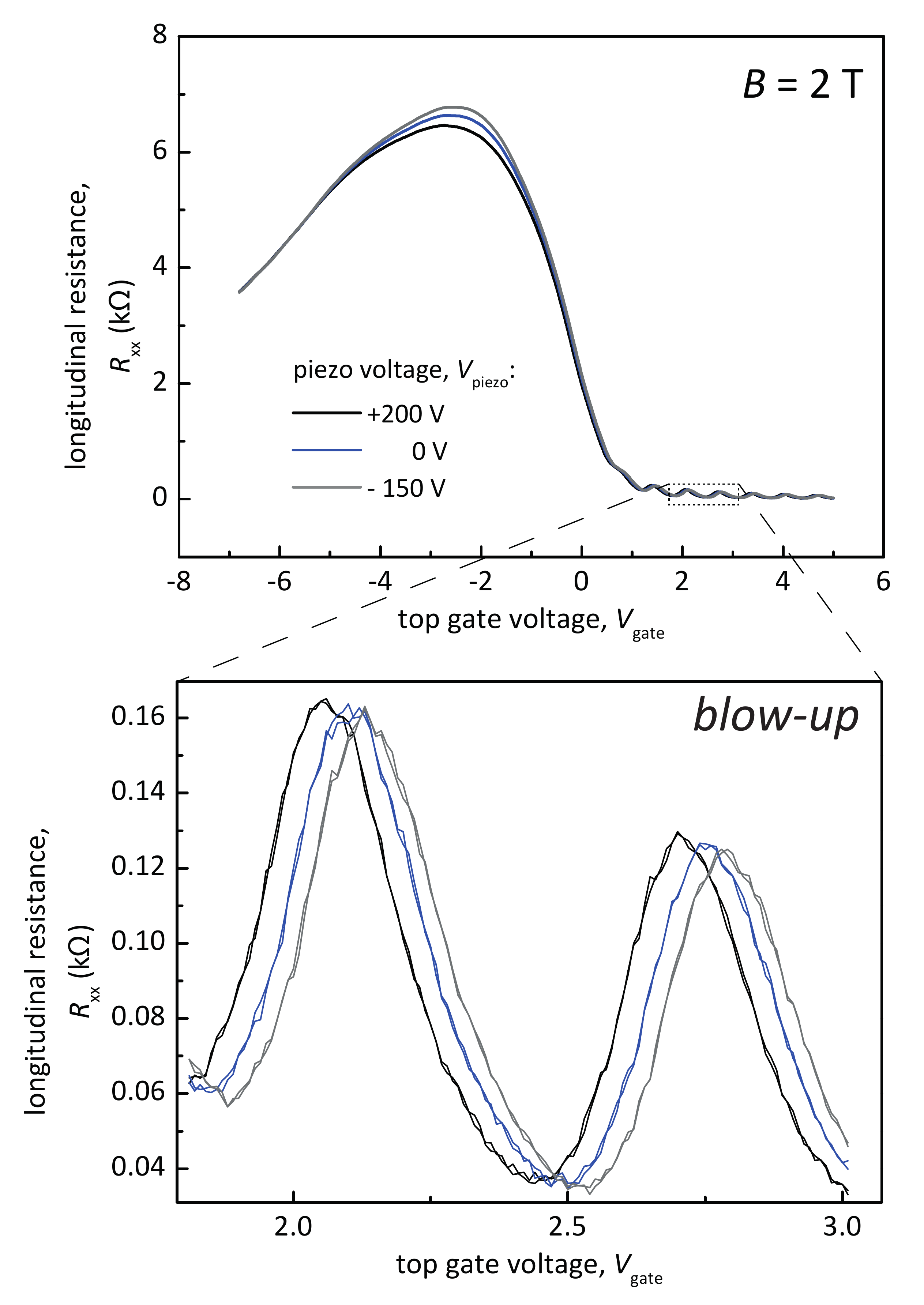}
\caption{\label{fig_appB1}{Additional exemplary measurements at 2 T (sample 4). At each magnetic field and each piezo voltage, we have performed two consecutive gate voltage sweeps to ensure the stability/reliability of the measurements. The bottom panel shows a blow-up of the marked area.  }}
\end{figure}

\newpage

\begin{figure}[h]
\includegraphics[width=8cm]{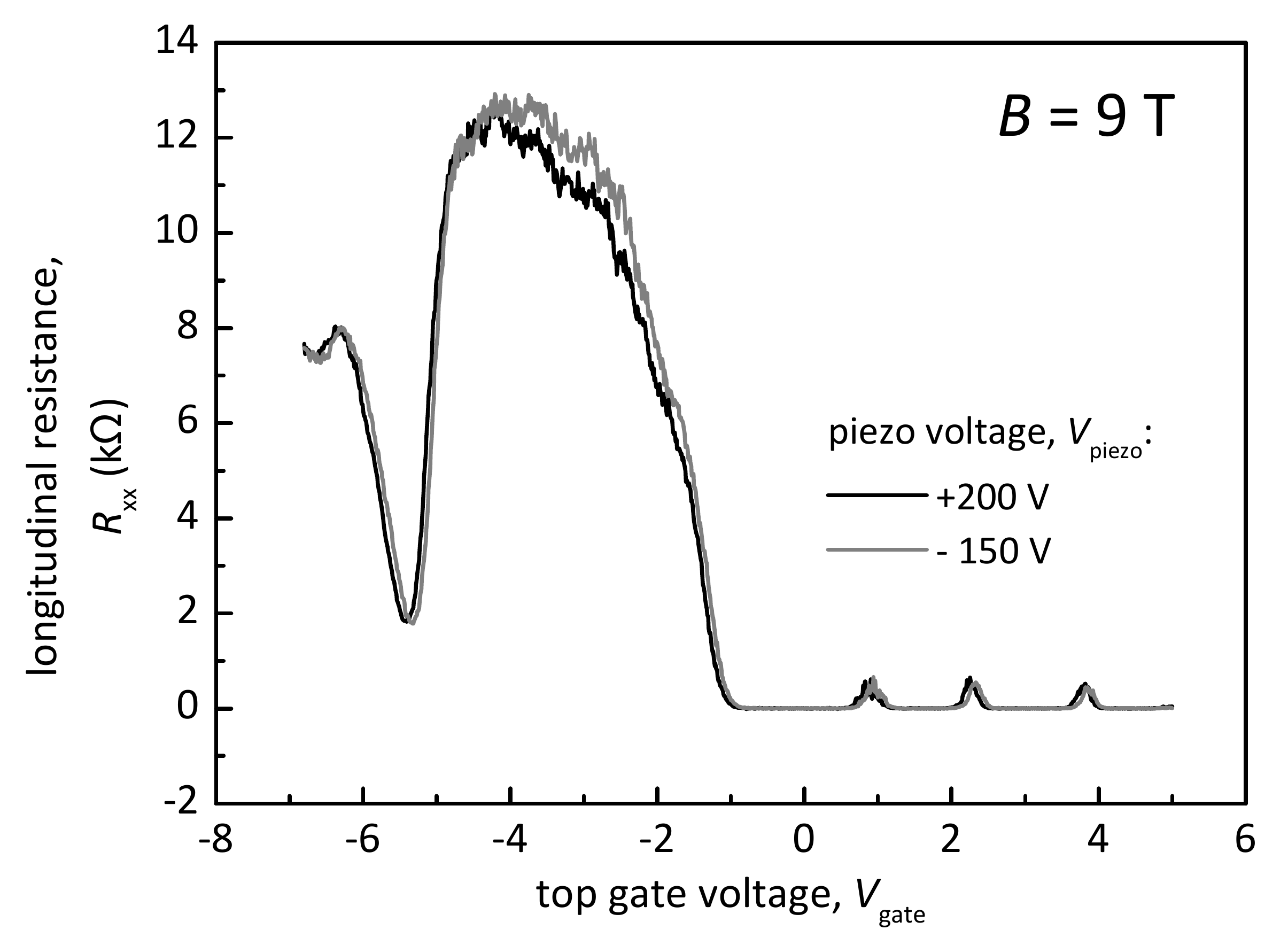}
\caption{\label{fig_appB2}{At large magnetic fields, the resistance often exhibits “noise” as shown for sample 4 at 9 T. This "noise" is responsible for the large error bars in Fig.\;\ref{fig3} (b).}}
\end{figure}

\newpage

\section[4]{\label{app_C} DOS TO DENSITY CHANGE CONVERSION }
To extract the density changes from the density of states (DOS), we start by determining the onsets of electron and hole conduction, located at the parabolic minimum of the electron-like band (denoted as $E_2$) and the parabolic maximum of the hole-like band (denoted as $E_1$). Next, we determine the effective mass of the electrons directly from the constant part of the density of states. Integration of the DOS yields the total density of electrons and holes. With the previously determined onsets $E_1$ and $E_2$ we can separate the total density into the respective electron and hole densities and thus determine the charge neutrality point (CNP). The latter determines the zero density point and the offset for the DOS integration. By doing these calculations for the DOS at all strain values, we determine how the charge carrier density changes due to strain.   

\section[5]{\label{app_D} BAND CALCULATION FOR 8 nm GaSb / 12 nm InAs}

\begin{figure}[h]
\includegraphics[width=8cm]{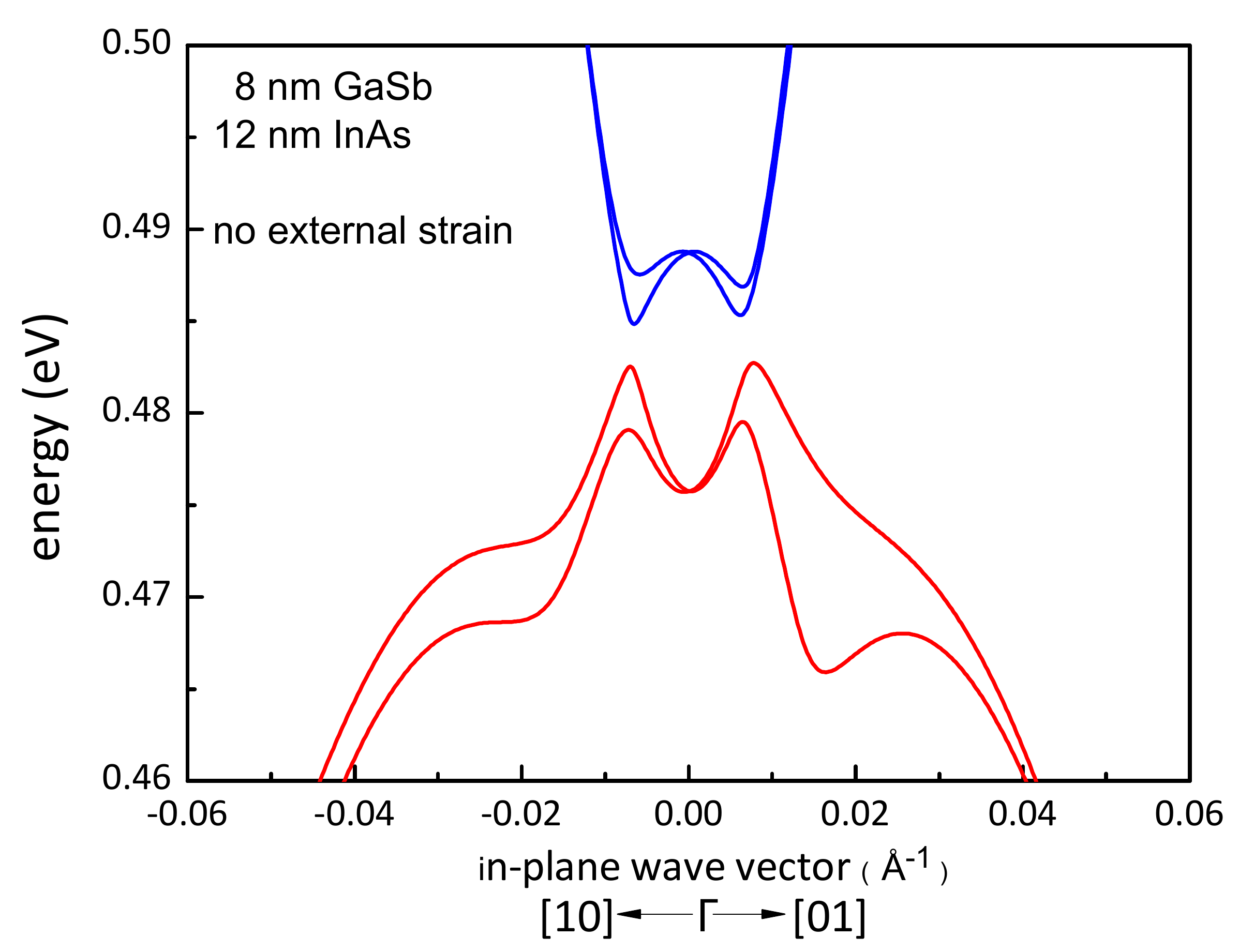}
\caption{\label{fig_appD}{Results of TB band calculations for an 8 nm GaSb / 12 nm InAs heterostructure using the same strain model as for our regular structure, i.e., the GaSb and the InAs quantum wells grow pseudomorphically with the same GaSb lattice constant. In contrast to the 8 nm / 15 nm structure, we now observe a clear mini band gap.}}
\end{figure}

\newpage

\end{appendix}

\newpage


\bibliography{tiemann_cite}
\bibliographystyle{unsrt}

\end{document}